# Periodic spatial variation of the electron-phonon interaction in epitaxial graphene on Ru(0001)

Andres Castellanos-Gomez[1,+,*], Gabino Rubio-Bollinger[1,2], Sara Barja[3], Manuela Garnica[3], Amadeo L. Vázquez de Parga[1,2,3], Rodolfo Miranda[1,2,3] and Nicolás Agraït[1,2,3].

We have performed low temperature scanning tunnelling spectroscopy (STS) measurements on graphene epitaxially grown on Ru(0001). An inelastic feature, related to the excitation of a vibrational breathing mode of the graphene lattice, was found at 360 meV. The change in the differential electrical conductance produced by this inelastic feature, which is associated with the electron-phonon interaction strength, varies spatially from one position to other of the graphene supercell. This inhomogeneity in the electronic properties of graphene on Ru(0001) results from local variations of the carbon – ruthenium interaction due to the lattice mismatch between the graphene and the Ru(0001) lattices.

The experimental realization of graphene [1] has boosted the research on this material, finding unique electronic properties [2-3]. The prospective use of graphene in optics, electronic devices [4] and chemical sensors [5] has also motivated a large number of studies devoted to develop different fabrication methods and characterization techniques. Among a variety of growth methods, epitaxial growth of graphene on transition metal substrates has been intensively studied because of its high efficiency and sample quality [6-8].

Nevertheless, the role of the substrate in the electronic properties of graphene devices can be very important [9] and it has to be studied in detail. For example, the extraction of photogenerated carriers in graphene photodetectors relies on spatial variations of the potential at the graphene/metal contacts. Often epitaxially grown graphene presents a variety of moiré patterns due to the lattice mismatch with the different transition metal substrates and a spatial modulation of the electronic structure, e.g. the local density of states of the graphene layer has been observed [10]. The Electron-Phonon Coupling (EPC) in graphene, which is responsible of a variety of properties from ballistic transport to excited-state dynamics, depends on the electronic density or doping level through the deformation potential [11]. A possible spatial modulation of the electron-phonon interaction, however, has not been studied in these systems.

[1] Departamento de Física de la Materia Condensada (C–III). Universidad Autónoma de Madrid, Campus de Cantoblanco, 28049 Madrid, Spain.
[2] Instituto Universitario de Ciencia de Materiales "Nicolás Cabrera". Campus de Cantoblanco, 28049 Madrid, Spain.
[3] Instituto Madrileño de Estudios Avanzados en Nanociencia IMDEA-Nanociencia, 28049 Madrid, Spain.
[+] Present address: Kavli Institute of Nanoscience, Delft University of Technology, Lorentzweg 1, 2628 CJ Delft (The Netherlands)
*E-mail: a.castellanosgomez@tudelft.nl

In this letter, we present low-temperature scanning tunnelling microscopy (STM) and inelastic electron tunnelling spectroscopy (IETS) measurements in monolayer graphene epitaxially grown on a Ru(0001) surface, which show a strong inelastic feature that can be attributed to the interaction between the tunnelling electrons and a breathing vibrational mode of the graphene layer. The intensity of this inelastic feature, which is proportional to the electron-phonon coupling strength and phonon density of states, varies spatially following the periodic Moiré pattern originated by the graphene/Ru(0001) lattice mismatch.

The samples were prepared in an ultra-high vacuum (UHV) chamber with a base pressure in the range of $10^{-11}$ Torr. The substrate is a single crystal of Ru exposing the (0001) surface, which was cleaned in UHV by ion sputtering and annealing to 1400 K. The graphene samples are grown under UHV conditions by thermal decomposition at 1000 K of ethylene molecules pre-adsorbed at 300 K on the sample surface. Depending on the amount of ethylene, nanometer-sized islands or a continuous, monolayer-thick graphene film that covers uniformly the Ru substrate over lateral distances larger than several microns can be prepared [8]. After the fabrication in UHV the sample was shortly exposed to air to be transferred to a $^3$He cryostat. The low reactivity of the graphene surface [12] makes possible this transfer without a noticeable degradation of the surface quality due to atmospheric contaminants. We use a homebuilt low temperature STM, operating at a base temperature of 300 mK under cryogenic vacuum, similar to the one described in ref.[13].

The STM tip has been first prepared *ex situ* by cutting a high purity (99.99 %) gold wire with scissors. Although these mechanically cut tips are rather usual in STM, the STM imaging and scanning tunnelling spectroscopy (STS) measurements are quite sensitive to the exact atomic configuration of the tip apex [14]. We have therefore employed a recently developed technique to prepare highly stable STM tips *in situ* at cryogenic temperatures [15]. Briefly, this method relies on local electric-field-induced deposition of material from the tip onto the studied surface. Subsequently, repeated indentations are gently performed onto the sputtered cluster to mechanically anneal the tip apex and thus to ensure the stability of the tip [15].

After the sample is transferred to the $^3$He cryostat, its quality is checked by measuring its STM topography in the constant current STM mode. The presence of monolayer graphene can be easily recognized in the STM images by a characteristic triangular array of bumps with an average separation of around 3 nm [8]. These bumps are originated by moiré pattern caused by the difference between the lattice parameter of graphene and Ruthenium surface. In fact, the graphene and Ru(0001) lattices are incommensurate, but 11 carbon honeycombs adjust almost exactly to 10 Ru-Ru interatomic distances forming the moiré pattern of bumps. Figure 1(a) shows a representative STM topograph in the cryogenic environment after *in situ* tip preparation. The measured topography is comparable to that acquired in samples that have been grown and measured in UHV [8]. Superimposed to the moiré pattern, Figure 1(a) also shows atomic resolution of the graphene lattice.

The lattice parameter mismatch also produces an inhomogeneous graphene-Ru interaction which modulates the electronic properties of graphene following the same moiré pattern as the topography [8]. Figure 1b shows scanning tunneling spectroscopy (STS) measurements recorded at different regions of the graphene/Ru (0001) unit cell (indicated by arrows in Figure 1(a)). The average differential conductance *vs.* voltage (d$I$/d$V$ *vs. V*) traces, obtained by numerical differentiation of current *vs.* voltage traces, at three selected locations of the Moiré pattern: a hill (A-region, where the graphene honeycomb is on top of a ruthenium atom) and two valley regions (hereafter B and C regions, where the graphene honeycomb is on a hcp hollow site and a fcc hollow site respectively). Each of the presented d$I$/d$V$ *vs. V* traces have been obtained by averaging 256 individual d$I$/d$V$ *vs. V* traces acquired at the specified locations.

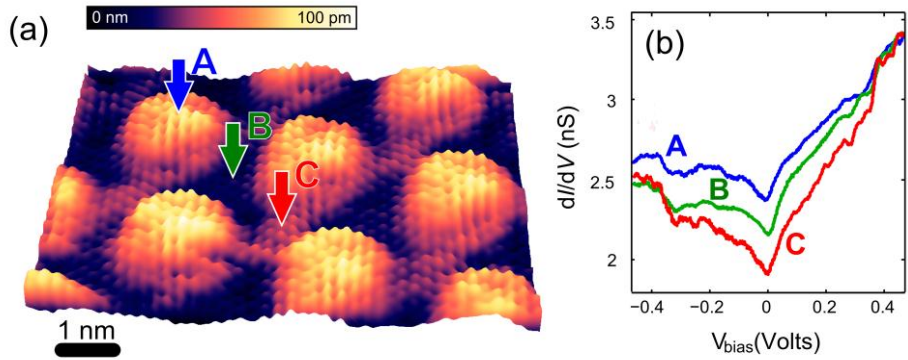

**Figure 1.** (a) High resolution STM topography of the graphene grown on Ru(0001) ($V_{sample}$ = 1 V; $I$ = 1.5 nA) at 300mK. (b) Averaged differential conductance as a function of the tip bias voltage measured at the regions A, B, C marked with arrows in (a). 256 traces measured at each region have been employed to obtain the averaged spectra.

The large drop in conductance at the Fermi level (Figure 1(b)) shows up in the IETS spectra as two peaks close to zero-bias. However, these peaks are not antisymmetric neither in height nor in voltage value and, thus, we rule out that they result from low energy phonon-excitations. A systematic study of graphite-graphite point contact suggests that this zero bias anomaly could be due to strong electron-electron Coulomb interactions or to quantum interference of the electrons 16. However, the study of this feature is out of the scope of the present work.

The traces show a clear asymmetry (although less marked in region A). They have a V-shape with a larger slope for positive tip biases. The conductance value at the Fermi level $E_F$, i.e. the zero-bias conductance value, is 20% larger on top of the hills (region A). Additionally, in the three regions there is a prominent increase of conductance (up to a 5-10%) around bias voltages of ± 360 mV. We assign this sudden change in conductance at symmetric positive and negative voltages to the inelastic scattering of the tunneling electrons due to the excitation of phonons in the lattice. In fact, when the tunneling electrons possess enough energy they can lose some energy by exciting vibrations of the graphene lattice, which results in an increased conductance of the tunneling junction and the opening of an extra tunneling channel 17.

Previous Raman spectra studies show that the most important vibrational modes in graphene are the G mode (with an energy of ~ 190 meV) and the 2D mode (also called G' mode, with an energy of ~ 340 meV), both with $E_{2g}$ symmetry 18. The energy of these phonons, however, can be slightly shifted by built-in strain in the graphene layer and electrostatic doping 19-20. Notice that both strain and doping effects are expected to be present on graphene monolayers grown on transition metals. For instance, graphene on Ru(0001) is highly doped as evidenced by a rather high density of states around zero bias (high d$I$/d$V$ in Figure 1(b)). Therefore it is not straightforward to attribute the observed features to a specific vibrational mode as both a second overtone of the G mode (a 2G mode, 380 meV) and the 2D mode (340 meV) are close to the IETS feature (360 meV), indicating that the observed IETS feature corresponds to the excitation of a second overtone of a phonon mode with symmetry $E_{2g}$ (a radial breathing mode of the honeycomb lattice).

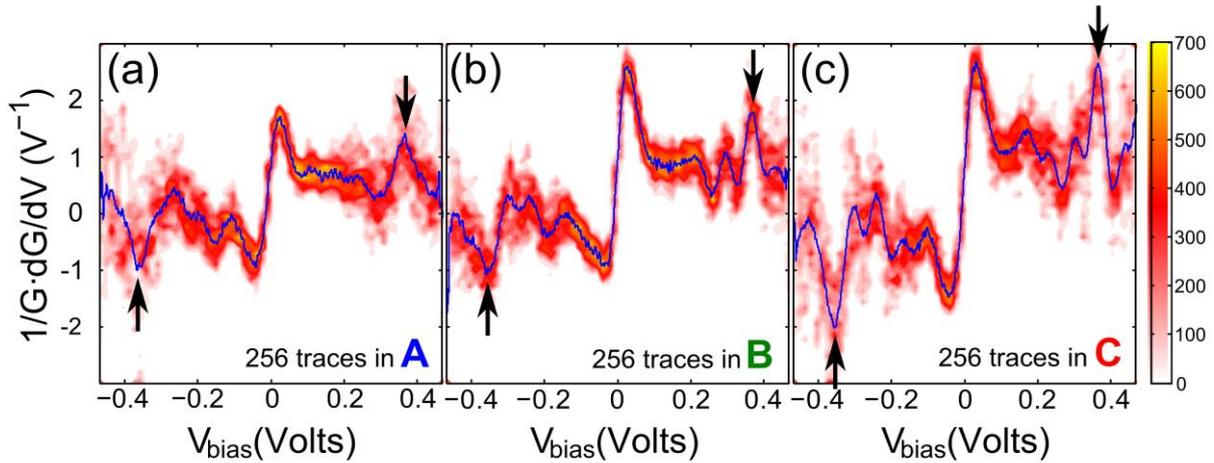

**Figure 2.** (a)-(c) Second derivative of the current $G^{-1} \cdot dG/dV$ as a function of the tip bias voltage built from the numerical differentiation of 256 $I/V(V)$ traces measured at the regions marked by arrows in Figure 1(a). The averaged $G^{-1} \cdot dG/dV$ trace is also plotted on top of the 2D histogram (solid blue line). The peaks at ± 360 mV indicate an inelastic tunneling process such as the excitation of phonons in the graphene lattice by electron-phonon interaction.

Studying the lattice vibrations by IETS presents a strong advantage with respect to Raman spectroscopy as the spatial resolution can be of the order of ~ 1 nm [21-22]. Figure 2 shows the IETS spectra measured at the three locations specified by the arrows in Figure 1(a). The second derivative of the current ($dI^2/dV^2$ *vs.* $V$) is plotted to facilitate the identification of the sudden changes in conductance due to the excitation of phonons. In this representation, the inelastic features appear as positive (negative) peaks at positive (negative) voltage. Moreover, dividing this representation by the differential conductance ($dI^2/dV^2$ / $dI/dV$ *vs.* $V$, hereafter called $G^{-1} \cdot dG/dV$ *vs* $V$ with $G$ standing for differential conductance), the height of these peaks is proportional to the electron-phonon interaction, i.e. the phonon density of states times the electron-phonon coupling strength [23]. The $G^{-1} \cdot dG/dV$ *vs* $V$ traces in Figure 2 are obtained by numerical differentiation of current *vs.* voltage traces. In order to extract statistical information from the IETS data, a whole set of $G^{-1} \cdot dG/dV$ *vs* $V$ traces (256 at each position in this case) is used to build a 2D histogram which allows one to easily visualize the most probable $G^{-1} \cdot dG/dV$ *vs* $V$ trace and its dispersion. To build these 2D histograms both the bias voltage and the $G^{-1} \cdot dG/dV$ axes are discretized into $N$ bins forming an $N$ by $N$ matrix (200 by 200 in our case) [24]. Each datapoint whose $G^{-1} \cdot dG/dV$ and $V$ values are inside a bin, adds one count to it. The number of counts in each bin is then represented with a color scale.

At -360 mV and +360 mV there are marked negative and positive peaks in the IETS spectra of similar absolute magnitude (see the black arrows in Figure 2), indicating the excitation of phonons at 360 meV by inelastic tunnelling processes. Interestingly, the observed magnitude of the IETS peaks, which is related to the electron-phonon interaction, varies substantially (> 50%) from the A-regions to the C-regions. This can be attributed to a difference in the graphene-Ru(0001) interaction, which is expected to be larger in the valleys (C-regions) than in the hills (A-regions) [25].

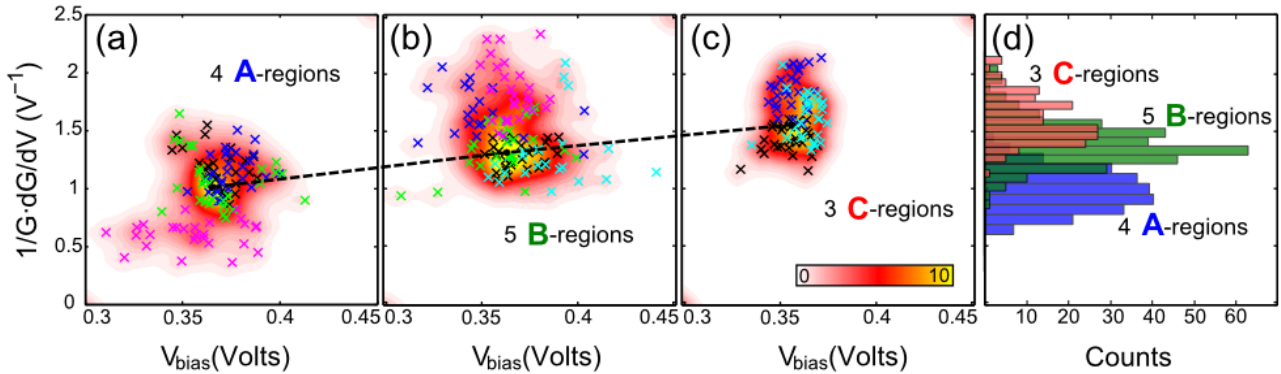

**Figure 3.** (a)-(c) 2D histograms of the voltage position and the intensity of the inelastic feature in the IETS spectra. Datapoints sharing the same color have been acquired in the same topographic region. A dashed line passing through the maximum of the 2D histograms has been included to point out that the intensity of the inelastic feature grows from A-regions to C-regions. (d) Comparison between the 1D histograms of the intensity of the IETS measured in regions A, B and C.

In order to systematically quantify this difference in electron-phonon interaction, we have analysed the IETS spectra measured at four A-regions, five B-regions and three C-regions. Figure 3 shows 2D histograms of the position and the magnitude of the peaks in the IETS spectra. We do not observe any variation of the phonon energy at different positions of the graphene lattice indicating that we are probing the same vibrational mode. The magnitude of the inelastic feature at ± 360 mV, on the other hand, shows a clear dependence on the position of the graphene supercell. The maximum density of datapoints in the 2D histogram obtained from the IETS spectra measured in C-regions occurs at a $G^{-1} \cdot dG/dV$ value 50% larger than the one of the A-regions showing larger electron-phonon interaction in the C-regions (valleys) than in the B-regions and A-regions (hills). We assign this spatial variation of the electron-phonon interaction to the inhomogeneous electronic structure that results from the different graphene/Ru(0001) interaction, in valleys and hills [25]. The excitation of the optical phonons, which depend on the relative displacement of the two sublattice atoms, are less probable on the hills. For instance, in A-regions (hills) the graphene honeycomb is on top of a ruthenium atom. This symmetric arrangement can hamper the excitation of breathing vibrational modes, in which the honeycomb atoms have to oscillate in anti-phase, resulting in a lower electron-phonon interaction in A-regions.

In conclusion, monolayer graphene epitaxially grown on Ru(0001) surface has been studied by low temperature scanning tunnelling microscopy and inelastic tunnelling spectroscopy at different positions of the moiré pattern arising from their lattice mismatch. A strong inelastic feature at 360meV, attributed to the excitation of phonons of a second overtone of a breathing vibrational mode of the graphene lattice has been observed to change its intensity with the same spatial periodicity of the moiré pattern. We found that the electron-phonon interaction, which can be of the utmost importance in the dynamics of charge carriers, is spatially modulated by the graphene/Ru(0001) interaction.

ACKNOWLEDGMENT

This work was supported by Comunidad de Madrid (Spain) through the program NANOBIOMAGNET (CAM s2009/MAT-1726), MINECO (Spain) through the programs MAT2011-25046 and CONSOLIDER-INGENIO-2010 CSD-2007-00010 and by the EU through the network "ELFOS" (FP7-ICT2009-6).